\documentclass[amsmath,amssymb,aps,pra,twocolumn]{revtex4-1}
\usepackage{graphicx}
\usepackage{dcolumn}
\usepackage{bm}
\usepackage{amsfonts} 
\usepackage{dsfont}
\usepackage{braket} 
\usepackage{hyperref} 
\usepackage{amsmath}
\usepackage{xcolor}

\begin{document}

\title{EPR steering in symmetrical Gaussian states}
\author{E. Benech$^{1,2}$, A. Auyuanet$^{1}$, A. Lezama$^{1}$}
\email{alezama@fing.edu.uy}
\affiliation{$^{1}$Instituto de F\'{\i}sica, Facultad de Ingenier\'{\i}a,
Universidad de la Rep\'{u}blica,\\ J. Herrera y Reissig 565, 11300
Montevideo, Uruguay\\
$^{2}$Institute for Quantum Optics and Quantum Information,
Austrian Academy of Sciences, A-1090 Vienna, Austria}
\date{\today}

\begin{abstract}
We have explored quantum Einstein-Podolsky-Rosen steering in symmetric two-mode Gaussian states using Gaussian and non-Gaussian measurements. 
For Gaussian measurements, we show that steering between the output modes of a symmetric beamsplitter is possible regardless of  purity when 
a threshold input-state quadrature variance compression is achieved. 
Using the non-Gaussian operators introduced in \citep{Ji2016} we show that non-Gaussian measurements can outperform Gaussian measurements for symmetrical states. We also analyze the possibility of asymmetric measurements setups made possible by non-Gaussian measurements and provide examples where such asymmetry is optimal for revealing steering. 
\end{abstract}

\maketitle
\section{Introduction}

Einstein-Podolsky-Rosen steering \citep{Einstein1935} (also known as EPR  entanglement) is a quantum correlation intermediate between entanglement and Bell-nonlocality. Unlike  these correlations, steering can be asymmetrical; one part (say Alice) may be able to steer the other part (Bob) while the reverse is not possible \citep{Quintino2015}. The term ``steering" was first used by E. Schr\"{o}dinger in 1935 \citep{Schrodinger1935}, to refer to the  ability of one system to influence the results of measurements carried on another when the two systems share an entangled state. While entanglement is a necessary condition for steering, not all entangled states are steerable \citep{Gisin1991,Wiseman07}.\\

In addition to its fundamental role in quantum theory \citep{Werner14}, steering has been identified as an important resource for quantum information processing.
Steering is required in two-party quantum protocols, such as quantum key distribution \citep{Branciard12,Cavalcanti13}, quantum teleportation with continuous variables \citep{CVteleportSteering} and randomness certification \citep{Passaro_2015}, for which entanglement certification is needed although one of the parties cannot be trusted. The study of steering was extended to the multipartite scenario \citep{genuinmultisteering,hierarchymultipartite} in connection with its applications to quantum communication networks \citep{3partisteeringopticalnet,multisteeringfreqcomb}.\\ 

A general definition of steering was formulated by Wiseman and coworkers \citep{Wiseman07}. If Alice and Bob share a global state, steering from Alice to Bob exists if the joint probability $P(a,b)$ of the outcomes $a$ and $b$ of measurements $A$ and $B$ carried respectively by Alice and Bob is not compatible with a local hidden state model where the probability of Bob's measurement outcomes are determined from a local quantum state through the rules of quantum mechanics.\\ 

This article is concerned with Gaussian continuous variables states. Gaussian states are ubiquitous in nature including such common states as the vacuum and thermal states. These examples refer to classical states however non-classical Gaussian states can also be produced via the application of squeezing operations. 

Our attention is specifically focused on a particular class of two-modes Gaussian states - symmetrical states - where Alice and Bob have access to the same amount of information. Symmetric states where one part can be steered by the other exists. However in this case steering is necessarily bidirectional; both parts are similarly able to \textit{steer} the other, provided the two perform the same measurements.\\ 

Within the vast ensemble of possible measurements one category, Gaussian measurements, plays an important role and has been extensively studied. Such measurements always amount to measuring field quadratures which can be performed via balanced homodyne detection. The conditions to be met by a two-mode Gaussian state in order to be steerable by Gaussian measurements are well known. They  will be reviewed below. \\

Two main questions are addressed in this paper. We first inquire whether non-Gaussian measurements can detect steerable states that are not steerable by Gaussian measurements. This question has previously received a positive answer for some examples of \emph{non-symmetrical} states. We show here that it is also the case for symmetric states. \\

We then explore the scenario where the two parts, Alice and Bob, have access to different measurement setups. More specifically we envision a scenario where Alice has a finite set of measurements at her disposal while Bob's measurements set is composed of a different number of observables. Such situation is asymmetric in spite of the state being symmetrical. 
We provide examples showing that asymmetric setups are able to reveal steering and  outperform Gaussian measurements with no loss of precision compared to larger symmetric setups.\\

The paper is organized as follows: In section \ref{estadossimetricos} we review the description of symmetrical two-mode Gaussian states and introduce a convenient parameterization of these states. In section \ref{steeringgaussiano}  the criterion for steering by Gaussian measurements is reminded and applied to the ensemble of symmetrical states. A threshold quadrature variance compression allowing steering regardless of state purity is derived. In section \ref{sec:non_gaussian} we remind the non-Gaussian observables introduced in \citep{Ji2016} and present several steering inequalities applicable to these observables. Section \ref{sec:results} presents the numerical evaluation of the steering inequalities and discusses the results. Concluding remarks are presented in section \ref{sec:conclusions}.

\section{Symmetrical two-mode Gaussian states}\label{estadossimetricos}

A quantum state is Gaussian if its Wigner phase-space representation is a Gaussian function. In consequence,  disregarding translations in phase space, all state properties are entirely defined by the knowledge of the covariance matrix (CM), i.e. the matrix of second order moments of the quadrature operators. 

A  two-modes symmetric Gaussian state (2MSGS) corresponds to  a $4 \times 4$ CM in the block form:
\begin{eqnarray}\label{standardformBlocks}
V_{S} &=&  \begin{pmatrix} A & C\\ 
C^T & A 
\end{pmatrix}
\end{eqnarray}
where the $2 \times 2$ blocks $A$ in the diagonal represent the local covariance matrices for Alice's and Bob's modes and block $C$ describes correlations between modes.\\ 

Using the fact that the quantum correlations between modes are not affected by local unitary transformations, it has been shown that the most general 2MSGS covariance matrix  can be brought through local symplectic transformations into the standard form \citep{Duan00,Adesso14,Marian08}:

\begin{eqnarray}\label{standardformu}
V_{S} &=&  \begin{pmatrix} pu & 0 & mu & 0\\ 
0 & \frac{p}{u} & 0 & \frac{n}{u}\\ 
mu & 0 & pu & 0\\ 
0 & \frac{n}{u} & 0 & \frac{p}{u}
\end{pmatrix}
\end{eqnarray}
with $p\geq 1$,  $p\geq \vert m \vert$, $p\geq \vert n \vert$, $(p^2-1)^2-2mn-p^2(m^2+n^2)+(mn)^2 \geq 0$ and $u >0$.\\

The parameter $u$ in \eqref{standardformu} can be arbitrarily modified (or made equal to one) through (symmetric) local squeezing. Therefore it does not affect the quantum correlation properties of symmetric states which are only dependent on the three relevant parameters $p, m, n$.

However,  parameters  $p, m, n$ are not necessarily the most practical for a systematic exploration of the  quantum correlation properties of symmetric states;  the  parameters are unbounded and not independent.\\

In this work we propose a new parameterization based on three \emph{independent} real parameters $\gamma$, $\mu$ and $\alpha$ subjected to the conditions:
\begin{equation}\label{condiciones}
(0 < \gamma \leq 1), (0 < \mu \leq 1), (0 \leq \alpha \leq 1)    
\end{equation}

The parameterization is physically motivated by the fact that mixing arbitrary Gaussian states (generally squeezed thermal states) in a symmetric beamsplitter results in a symmetric two-mode state. 

Specifically, we consider the 2MSGS obtained by sending through a symmetric BS the state described by the covariance matrix:
\begin{eqnarray}\label{estadoparametrizado}
V_{\gamma,\mu,\alpha} 
& =& Diag \left[ \frac{\gamma^{-1}}{\mu},\frac{\gamma}{\mu},\left(\frac{\gamma}{\mu}  \right)^{\alpha},\left(\frac{\gamma^{-1}}{\mu}  \right)^{\alpha}\right]
\end{eqnarray}

The state described by \eqref{estadoparametrizado} represents two squeezed thermal states incident on the two input ports of the BS. It is always  a physical state if the conditions \eqref{condiciones} are satisfied. \\

It can be shown (details are given in Appendix) that for any symmetric Gaussian state corresponding to parameters $p, m, n$, a set of four non-negative real numbers $\gamma, \mu, \alpha, u$ can be found for which the passage of state \eqref{estadoparametrizado} through a symmetric BS results in state \eqref{standardformu}. The obtained value of $\mu$ verifies $0 < \mu \leq 1$ and with no loss of generality the conditions $0 < \gamma \leq 1$ and $0\leq \alpha \leq 1$ can be imposed.\\

\begin{figure}[ht]
\centering
\includegraphics[width=1\linewidth]{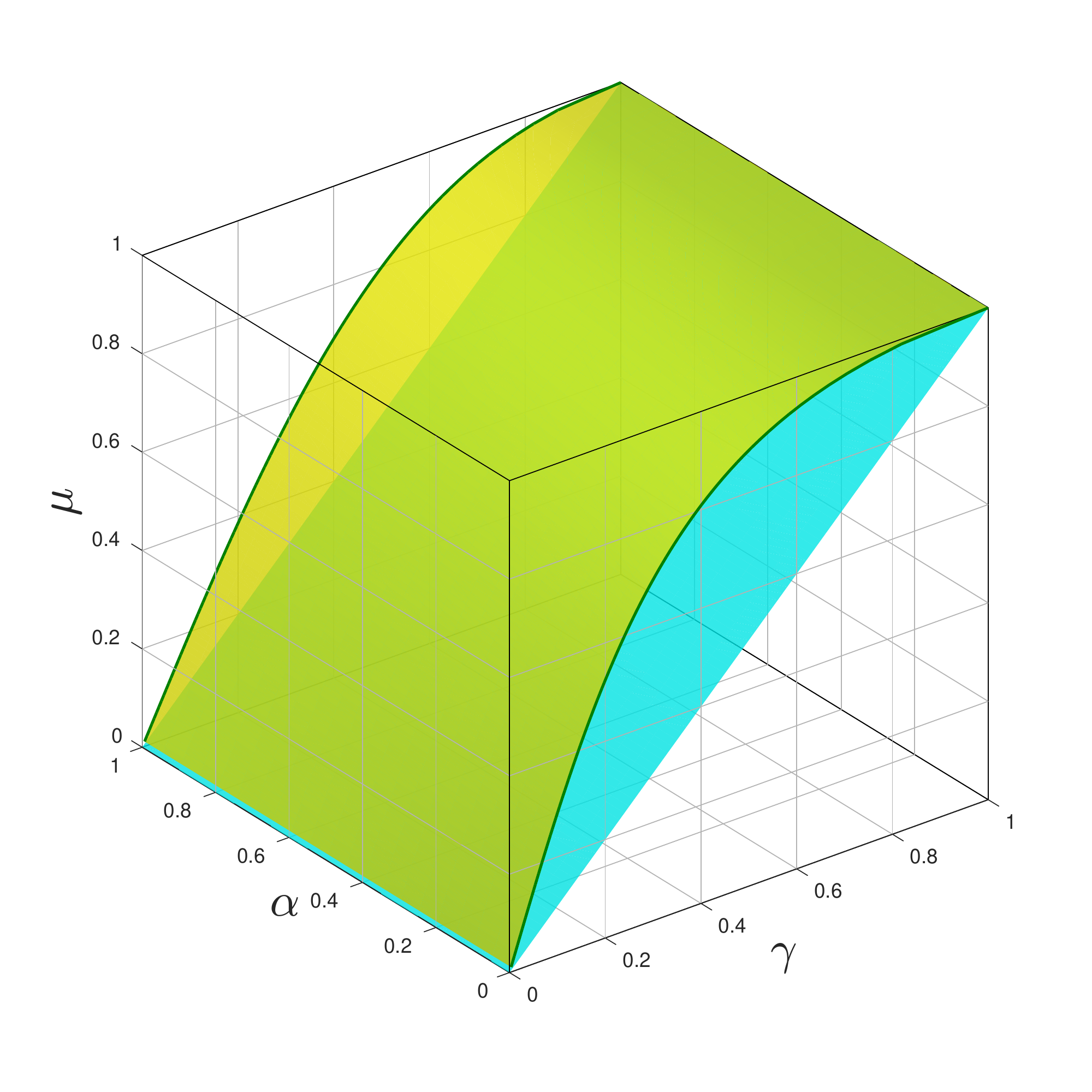}
\caption{\label{cubo} Unit cubic volume representing all symmetric two-modes Gaussian states. States above the diagonal plane are entangled. The states above the yellow surface are steerable by Gaussian measurements.} 
\end{figure}

We therefore conclude that, as far as quantum correlations are concerned, all two-mode symmetrical Gaussian states can be represented by the points of the the $\gamma, \mu, \alpha$ parameter space contained in a unit-volume cube (Fig. \ref{cubo}).\\

The convenience of our parameterization for the study of quantum correlations in 2MSGS can be illustrated by the simplicity of the condition for entanglement:  $\gamma < \mu$. A violation of this condition results in classical states entering the beamsplitter and consequently a separable output state \citep{Kim2002,Wolf2003}. The condition is also sufficient (see remark in Appendix).\\

Our parameters choice is also motivated by the fact that it allows continuous variation of the 2MSGS between two limiting cases of interest. When $\alpha =1$ the resulting state corresponds to a symmetric two-mode squeezed thermal state (2MSTS) with squeezing coefficient $\gamma$ and purity $\mu_{Tot}= \mu^{2}$. When $\mu=1$ the output state corresponds to the pure two-mode squeezed vacuum  state (2MSV). On the other hand, $\alpha =0$ corresponds to a situation frequently encountered in experiments where a squeezed thermal state is incident on one of the input ports of a BS while vacuum is incident on the other. We refer to the corresponding two-mode state as a squeezed-thermal and vacuum state (STVS).\\

2MSTS have been thoroughly studied (see for example \citep{Laurat_2005, Xiangsqueezed, MarianSymmetricG,CHEN2003191}). Recently they have been considered in the context of  quantum metrology \citep{Li_2016} and the extraction of quantum work \citep{Cuzminschi2021}. The properties of STVS have been less explored in spite of being easily produced in experiments.

\section{Steering by Gaussian measurements}\label{steeringgaussiano}

The condition for steering by Gaussian measurements in two-mode Gaussian states was determined by Kogias and co-workers \citep{Kogias15}. Steering by Alice of Bob's state occurs iff:
\begin{equation}\label{steering_gaussiano}
\det(A) > {\det(V)}
\end{equation}
here A refers to the CM for Alice mode and V is the total CM of the two-mode system. The purity $\eta$ of a state  is related to its covariance matrix $V$ by $\eta = 1/\sqrt{\det(V)}$. In consequence, Eq. \eqref{steering_gaussiano} implies that in the case of two-mode Gaussian states \textit{and Gaussian measurements}, the sufficient and necessary condition for Alice to be able to steer Bob's system is that her reduced state is less pure than the whole state.\\

For the state obtained from \eqref{estadoparametrizado} after passage through the BS, condition \eqref{steering_gaussiano} becomes:

\begin{eqnarray}
\mu^{2(1+\alpha)}X^{2}-(4-\mu^{2\alpha}-\mu^{2}) X+1 &\geq & 0\label{desigualdad} 
\end{eqnarray}
where $X\equiv (\gamma/\mu)^{1+\alpha}$.

\begin{figure}[ht]
\begin{center}
\includegraphics[width=1\linewidth]{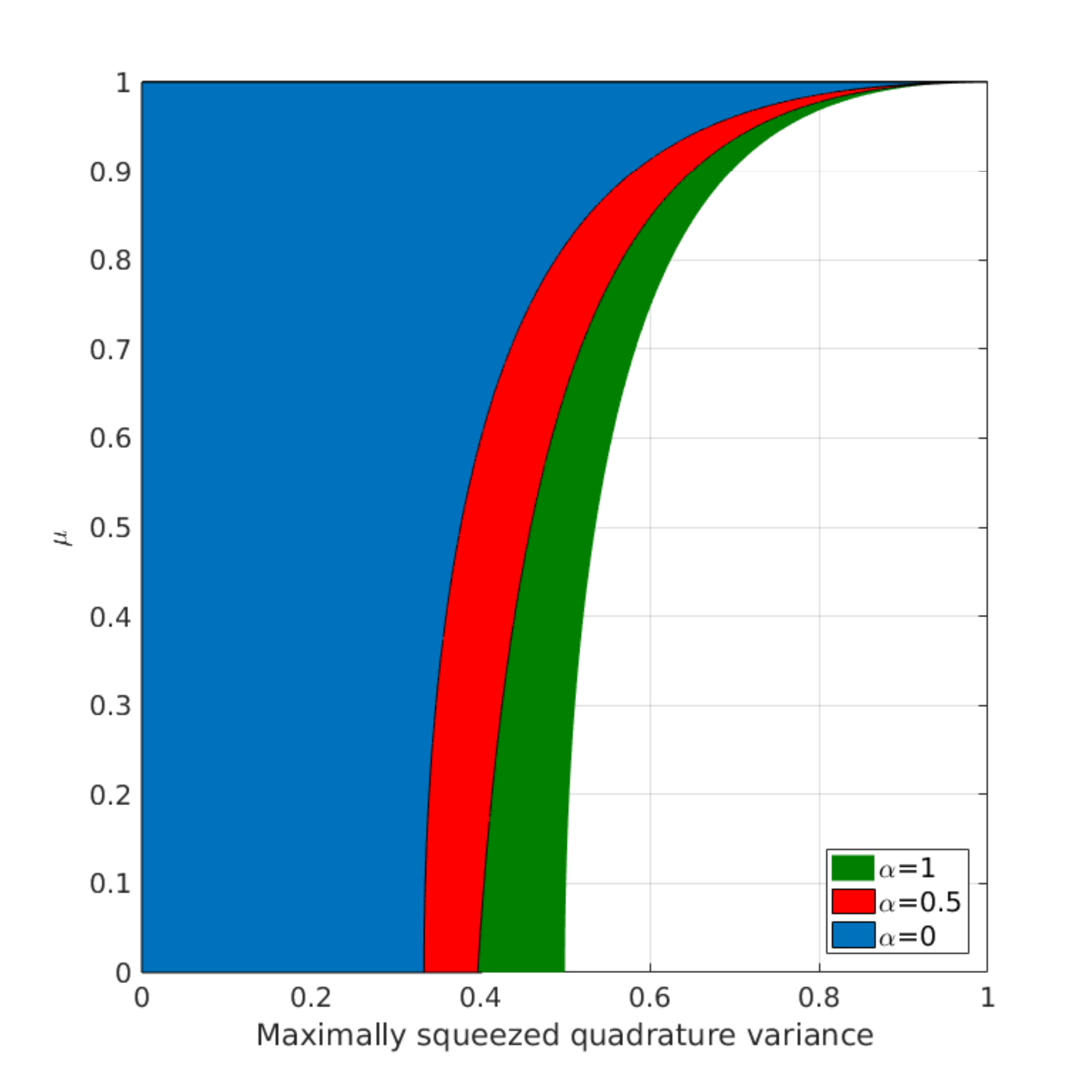}
\caption{States steerable by Gaussian measurements. The filled areas (overlapping) correspond to steerable symmetrical two-mode Gaussian states obtained by sending the state defined by CM \eqref{estadoparametrizado} through a $50 \%$ transmission beamsplitter.
\label{estadoestirables}}
\end{center}
\end{figure}

The surface separating steerable and non-steerable states by Gaussian measurements is obtained by solving \eqref{desigualdad} for equality. It is represented in Fig. \ref{cubo}.\\

The states steerable by Gaussian measurements for three different values of the parameter $\alpha$ are represented in Fig. \ref{estadoestirables} where the horizontal axis corresponds to  the maximally squeezed quadrature variance $v \equiv \gamma/\mu $ in \eqref{estadoparametrizado}. Interestingly, there exists a threshold value $v_{th}$ of the maximally squeezed quadrature variance of the BS input-state below which steering by Gaussian measurements between output modes is possible regardless of purity. In the general case ($\alpha > 0$),  $v_{th}=(1/4)^{\frac{1}{1+\alpha}}$. For $\alpha =0$ (STVS) $v_{th} = 1/3$. The existence of such a threshold, not previously identified, is relevant to experiments.

\section{Steering by non-Gaussian measurements}\label{sec:non_gaussian} 

So far we have been concerned only with Gaussian measurements. One can inquire whether non-Gaussian measurements can reveal steerable states that are not identified through GM.\\ 

It has been conjectured that GM are optimal for detecting steering in Gaussian states \citep{Kogias_15}. Prior to our work this conjecture has not been disproved in the case of symmetrical states such as two-mode squeezed vacuum states (TMSV). However, it was shown \citep{Tatham14,Ji16}  that when asymmetric losses act upon a TMSV state, the conjecture is no longer true. In this work we demonstrate that non-Gaussian measurements can outperform Gaussian measurements also in the case of symmetric states. 
In addition, non-Gaussian measurements allow addressing the scenario where Alice and Bob perform measurements chosen from different sets of observables (different setups). This is a rich scenario which has been seldom addressed for continuous variables states. It is not possible for GM which always amount to quadrature measurements by both parties.\\

A steering criterion for arbitrary non-Gaussian measurements is generally not available and must be established on a case by case basis. In general, when such a criterion can be derived, it is not directly expressed in terms of the two-mode CM.

An experimentally useful criteria for the EPR paradox was proposed by Reid \citep{Reid89}.

Later, Cavalcanti and Reid have shown that steering inequalities can be derived for experimentally accessible observables constrained by multiplicative or additive uncertainty relations \citep{Cavalcanti2007}. A powerful method for the derivation of steering criteria for arbitrary measurement sets relying on experimentally observable quantities was established in \citep{Cavalcanti09}. \\

In our work we have used the set of non-Gaussian observables introduced by Ji and coworkers \citep{Ji2016}. The set of observables available to Alice is designated as $\{A^{(n)}_{i}\}$ where the superscript ${(n)}$ identifies the whole set and the subscript $i$ one of its members. 

The set $\{A^{(n)}_{i}\}\equiv\{\lambda_{k},\lambda_{kl}^{\pm}\}$ contains $n^2$ projective measurements defined as: 
\begin{subequations}\label{operadoresdeJi}
\begin{eqnarray}
\lambda_k &=& \vert k \rangle \langle k \vert,\\
\lambda_{kl}^{+} &=& \frac{\vert k \rangle \langle l \vert +\vert l \rangle \langle k \vert}{\sqrt{2}}\qquad (k<l),\\
\lambda_{kl}^{-} &=& \frac{\vert k \rangle \langle l \vert -\vert l \rangle \langle k \vert}{\sqrt{2}i}\qquad (k<l),
\end{eqnarray}
\end{subequations}
where $k,l=0,1,\cdots ,n-1$ and $\vert k \rangle$ and $\vert l \rangle$ are Fock states. These observables  are orthogonal projectors into Fock-state-pair  subspaces. Similar definitions apply to Bob's observables set $\{B^{(n^{\prime})}_{j}\}$.\\
  
From measurements of the above observables a correlation matrix can be defined:
\begin{eqnarray}\label{matrizdecorrelacion}
C^{nn^{\prime}}_{i,j} &=& \langle A^{(n)}_i\otimes B^{(n^{\prime})}_j \rangle -\langle A^{(n)}_i \rangle \langle B^{(n^{\prime})}_j \rangle
\end{eqnarray}

The observables on Bob's mode must satisfy the  uncertainty relation \citep{Ji16}
\begin{eqnarray}
\sum_{j=1}^{{n^{\prime}}^2} \Delta^2(B_j^{(n^{\prime})})\geq (n^{\prime}-1)\langle \mathds{1}_B^{(n^{\prime})} \rangle\label{incertidumbredeJi}
\end{eqnarray}
where $\Delta^2(X)= \braket{X^2}-\braket{X}^2$ is the variance of the observable $X$ outcomes and $\mathds{1}_B^{(n^{\prime})} \equiv \sum_{k=0}^{k=(n^{\prime}-1)} \ket{k}\bra{k}$ refers to the identity operator truncated to Fock states with photon number smaller than $n^{\prime}$. A similar uncertainty relation holds for Alice's observables. \\

Applying the result in \citep{Cavalcanti09} to the uncertainty relation \eqref{incertidumbredeJi} one can show that if  Bob's state is \emph{not} steerable by Alice measurements then:

\begin{eqnarray}
\min_{\lbrace A^{(n)}_j \rbrace} \sum_{j=1}^{j={n^{\prime}}^2} \Delta_{inf}^2(B_j^{(n^{\prime})}\vert A^{(n)}_j) \geq (n^{\prime}-1)\langle \mathds{1}_B^{(n^{\prime})} \rangle \label{cavalcanti_est}
\end{eqnarray}
Here $\Delta_{inf}^2(B_j^{(n^{\prime})}\vert A^{n}_j)$ refers to the inferred variance of the output of measurement $B_j^{(n^{\prime})}$ in view of the output of Alice's measurement of observable $A^{n}_j$.  The inferred variance of observable $B$ is defined as:

\begin{eqnarray}\label{varinf}
\Delta^{2}_{inf}(B\vert A) &=&
\sum_{a,b}P(a,b)(b-b_{est}(a))^{2}
\end{eqnarray}
where $a$ and $b$ refer to the outcomes of Alice and Bob realizing measurements $A$ and $B$ respectively. $P(a)$ is the probability of outcome $a$ and $P(a,b)$ the joint outcome probability. $b_{est}(a)$ designates an estimate of the output $b$ of the measure by Bob of observable $B$ given that the output of the measurement of observable $A$ by Alice was $a$. $b_{est}(a)$ is an arbitrary function of $a$ generally chosen to satisfy: $\sum_{b}P(b)b=\sum_{a}P(a)b_{est}(a)$.

There is no general procedure for constructing $b_{est}(a)$. A bad estimation will result in an increase of the inferred variance resulting in a weaker inequality \eqref{cavalcanti_est}. Examples of two convenient choices for $b_{est}(a)$ are provided below.\\

It is worth stressing at this point that the sets of observables ${\lbrace A^{(n)}_j \rbrace}$ and ${\lbrace B^{(n)}_j \rbrace}$ appearing in the LHS of inequality \eqref{cavalcanti_est} can be different, sub-index $j$ is being used to identify a pair which do not need to be equal. Also the observables contained in the set ${\lbrace A^{(n)}_j \rbrace}$ that minimize the LHS inequality \eqref{cavalcanti_est} are not required to be all different.  Consequently inequality \eqref{cavalcanti_est} can be used to test steerability with asymmetric (different) measuring setups for Alice and Bob.\\

From inequality \eqref{cavalcanti_est} Ji et al. \citep{Ji16} have derived a steering inequality involving the trace norm of the correlation matrix  $C^{nn^{\prime}}$. Steering from A to B is possible if:

\begin{align}
 &\Vert C^{nn^{\prime}} \Vert_{tr} > \nonumber\\ 
&\sqrt{\left( n\langle \mathds{1}_n^{A}\rangle  -\sum_{j=1}^{n^{2}}\langle A_j^{(n)}\rangle^2 \right)\left( \langle \mathds{1}_{n^{\prime}}^{B}\rangle  -\sum_{j=1}^{{n^{\prime}}^{2}}\langle B_j^{(n^{\prime})}\rangle^2 \right)}, \label{condicionJi}
\end{align}
where $\Vert \bullet \Vert_{tr}$ designates the trace norm.\\

The starting point of our analysis is also inequality \eqref{cavalcanti_est} whose violation implies steering. We have used two different choices of the function $b_{est,j}$ which gives the estimated value of the output $b$ of measurement $B_j$ by Bob given that Alice has obtained output $a$ from her measurement of $A_j$.\\

Our first choice is:
\begin{eqnarray}
b_{est,j}(a)=\langle B_j \rangle_{a}\equiv \langle  \mathds{1}_{A}\otimes B_j \rangle_{a},
\end{eqnarray}
where the subscript $a$ indicates that the mean value is taken on the state of the system after measurement of $A_j\otimes\mathds{1}_{B}$ with output $a$. This choice of $b_{est}(a)$ ensures that the inferred variance $\Delta^{2}_{inf}(B_j\vert A_j)$ attains its minimum value $\Delta^{2}_{min}(B_j\vert A_j)$ \citep{Reid_2009,Cavalcanti09} and can consequently be considered as optimum.
With this choice inequality \eqref{cavalcanti_est} becomes:

\begin{eqnarray}
\min_{\lbrace A^{(n)}_i \rbrace} \sum_{j=1}^{{n^{\prime}}^2} \Delta_{min}^2(B_j^{(n^{\prime})}\vert A^{(n)}_i) \geq (n^{\prime}-1)\langle \mathds{1}_B^{(n^{\prime})} \rangle \label{laminima}
\end{eqnarray}

It should be noted that in order to evaluate $\langle B_j \rangle_{a}$ the knowledge of the total two-mode state density matrix is required, which cannot be obtained from local measurements only.  In consequence, the identification of steerable states obtained by violation of inequality \eqref{laminima} with this estimate should be considered as a theoretical limit.\\ 

Our second choice consists in the linear estimate (also considered in \citep{Cavalcanti09}, and shown on \citep{Reid_2009}):
\begin{eqnarray}
b_{lin\_est,j}(a)=-g_j (a-\langle A_j \rangle) +\langle B_j \rangle,
\end{eqnarray}
where $g_j$ is a real number. The best choice for $g_j$ in order to minimize the inferred variance $\Delta_{inf}^2(B_j\vert A_j)$ is $g_j=- [ \langle B_jA_j \rangle-\langle B_j\rangle \langle A_j \rangle] [\langle A_j^2 \rangle -\langle A_j \rangle^2]^{-1}$. With this choice inequality \eqref{cavalcanti_est} becomes after some manipulation:

\begin{align}
\max_{\lbrace A^{(n)}_j \rbrace} \left\lbrace \sum_{j=1}^{{n^{\prime}}^2} \frac{ [ \langle B_jA_j \rangle - \langle B_j\rangle \langle A_j \rangle]^2}{ \langle A_j^2 \rangle -\langle A_j \rangle^2}  \right\rbrace &\nonumber\\ 
\leq  \langle \mathds{1}_B^{(n^{\prime})} \rangle-\sum_{j=1}^{{n^{\prime}}^2} \langle B_j\rangle^2 &\label{ladeEuge}.
\end{align}

The results of the evaluation of inequalities for symmetric Gaussian states is presented next.

\section{Results}\label{sec:results}

We have numerically computed the two sides of inequalities \eqref{condicionJi}, \eqref{laminima} and \eqref{ladeEuge} for the symmetric states corresponding to parameters $\gamma, \mu, \alpha$. The calculation requires the knowledge of the state density matrix in the Fock basis. We have calculated the coefficients of the density matrix truncated to a maximum photon number ($N=19$) using multivariate Hermite polynomials \citep{Tatham14,Ji16}. It was checked that the truncation of the density matrix does not affect the results presented below which concern states for which the values of $\gamma$ and $\mu$ are relatively large (above $0.2$). As these parameters are further reduced, density matrix coefficients left aside by the truncation are expected to play an increasing role. Density matrix coefficients were calculated over a grid of 441 equally spaced states within the rectangle in the ($\gamma, \mu$) plane  presented in each figure. The contour lines corresponding to equality were obtained by extrapolation using Matlab.\\

\begin{figure}[htb]
\centering
\includegraphics[width=1.05
\linewidth]{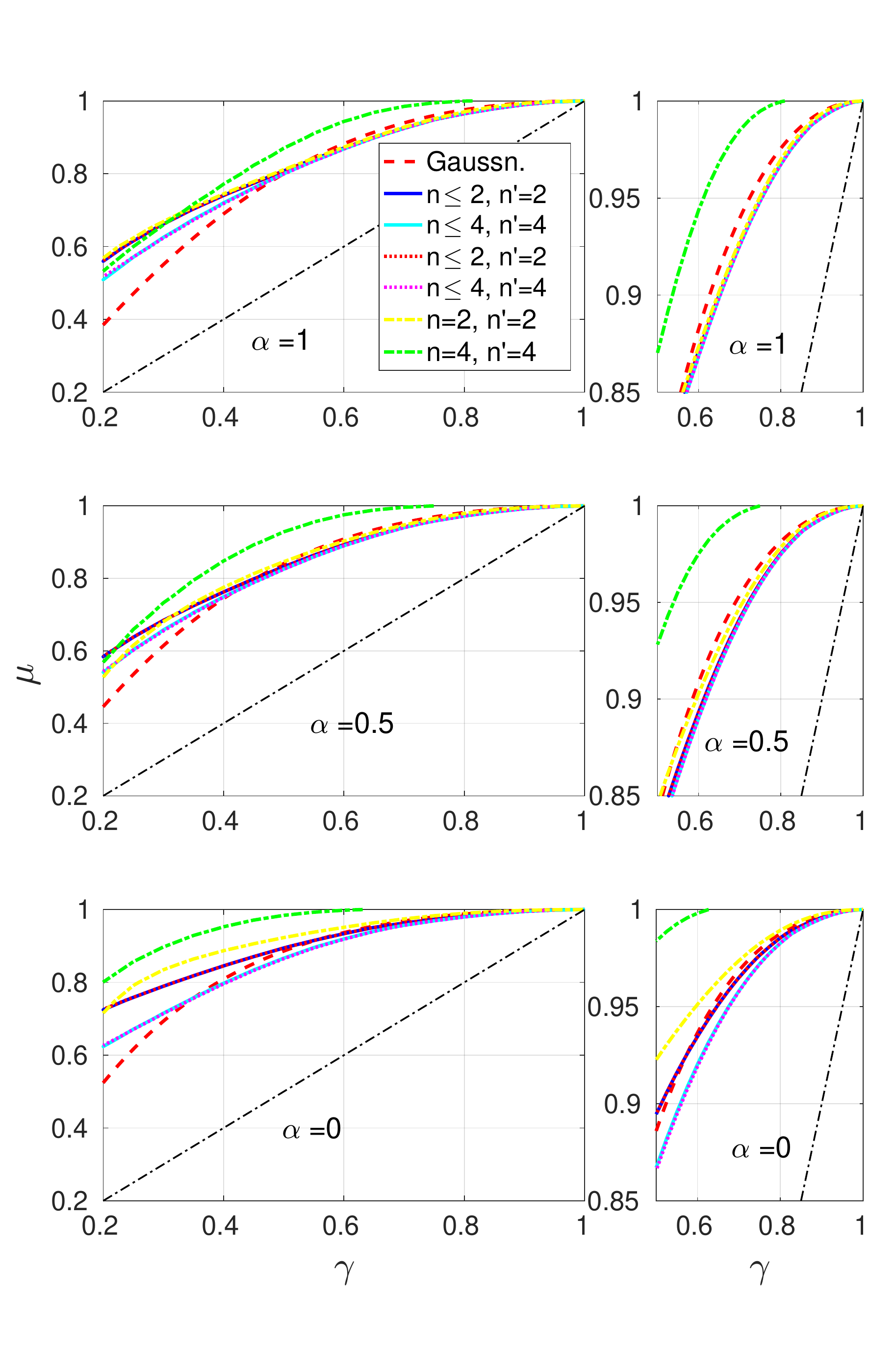}
\caption{\label{JEnn} Left column: steerable states according to  inequalities \eqref{ladeEuge} (solid), \eqref{laminima} (dotted), and \eqref{condicionJi} (dash-dotted) for different values of $\alpha$. Steering is possible in the region above the  limiting lines. Dashed line: steerability limit for  Gaussian measurements. States under the black dash-dotted line are separable. Right column: partial enlarged view.} 
\end{figure}

\subsection{Symmetric setups}\label{sec:symmetric}

Figure \ref{JEnn}  present the results obtained for symmetric ensembles of observables  ${\lbrace A^{(n)}_j \rbrace}$ and ${\lbrace B^{(n)}_j \rbrace}$ corresponding to $n=1$ and $n=4$. The corresponding limit for Gaussian measurements is indicated in each figure. Steering from Alice to Bob is possible when either inequality \eqref{condicionJi} (dotted) or \eqref{ladeEuge} (solid) is violated. Steerable states correspond to the area  above the limiting curves. It is worth reminding here that the minimization in \eqref{cavalcanti_est}  does not require that all operators in Alices's set ${\lbrace A^{(n)}_j \rbrace}$ are used. It could occur that the minimum is reached using only operators contained in a smaller set corresponding to a lower value of $n$. Such possibility is indicated by the inequality in the figure legend and will be discussed in more detail in the next section.\\

The results in Fig. \ref{JEnn} show that for some symmetric states, the non-Gaussian observables considered in this work can outperform Gaussian measurements in revealing steerable states for all values of the parameter $\alpha$. This includes the case $\alpha=1$ corresponding to two-mode squeezed thermal states. 

As expected, violation of inequality \eqref{laminima} which uses the minimum inferred variance is stronger than violation of \eqref{ladeEuge} based on linear estimates with different parameters for each observable. Inequality \eqref{ladeEuge} is in turn stronger than violation of \eqref{condicionJi} based on a single-parameter linear estimate.\\

It is interesting to signal that on the scale of Fig. \ref{JEnn} the limiting curves corresponding to inequality \eqref{ladeEuge} which is based on locally accessible quantities is barely distinguishable from the theoretical limit provided by inequality \eqref{laminima}. Also, considering that all inequalities were derived from \eqref{cavalcanti_est} it is remarkable that inequality \eqref{condicionJi} behaves poorly as $n=n^{\prime}$ is increased. This can be traced to the fact that a single coefficient $g$ is used for the linear estimate of all observables and that the use of the trace norm  requires different observables on Alice's set \citep{Ji2016}.\\

In the regions of Fig. \ref{JEnn} corresponding to small values of $\gamma$ and $\mu$, Gaussian measurements remain more effective for revealing steering than the considered non-Gaussian observables. This behavior is expected since for such states a large amount of the information is contained in density matrix coefficients corresponding to large Fock number states which are not addressed by the non-Gaussian observables. On the other hand, the steering criterion for Gaussian observables makes use of the covariance matrix which contains all the information regarding the state. 
\subsection{Asymmetric setups}\label{sec:assymmetric}

The use of non-Gaussian observables allow the consideration of scenarios where the setups available to Alice and Bob corresponding to the sets ${\lbrace A^{(n)}_i \rbrace}$ and ${\lbrace B^{(n^{\prime})}_j \rbrace}$ respectively are different.
The inequalities \eqref{ladeEuge} and \eqref{laminima} are well suited for the consideration of asymmetric setups. On the other hand,  inequality \eqref{condicionJi}  requires that $n\geq n^{\prime}$ and is not optimum for $n\neq n^{\prime}$ \citep{Ji2016}.\\

\begin{figure}[htb]
\includegraphics[width=1\linewidth]{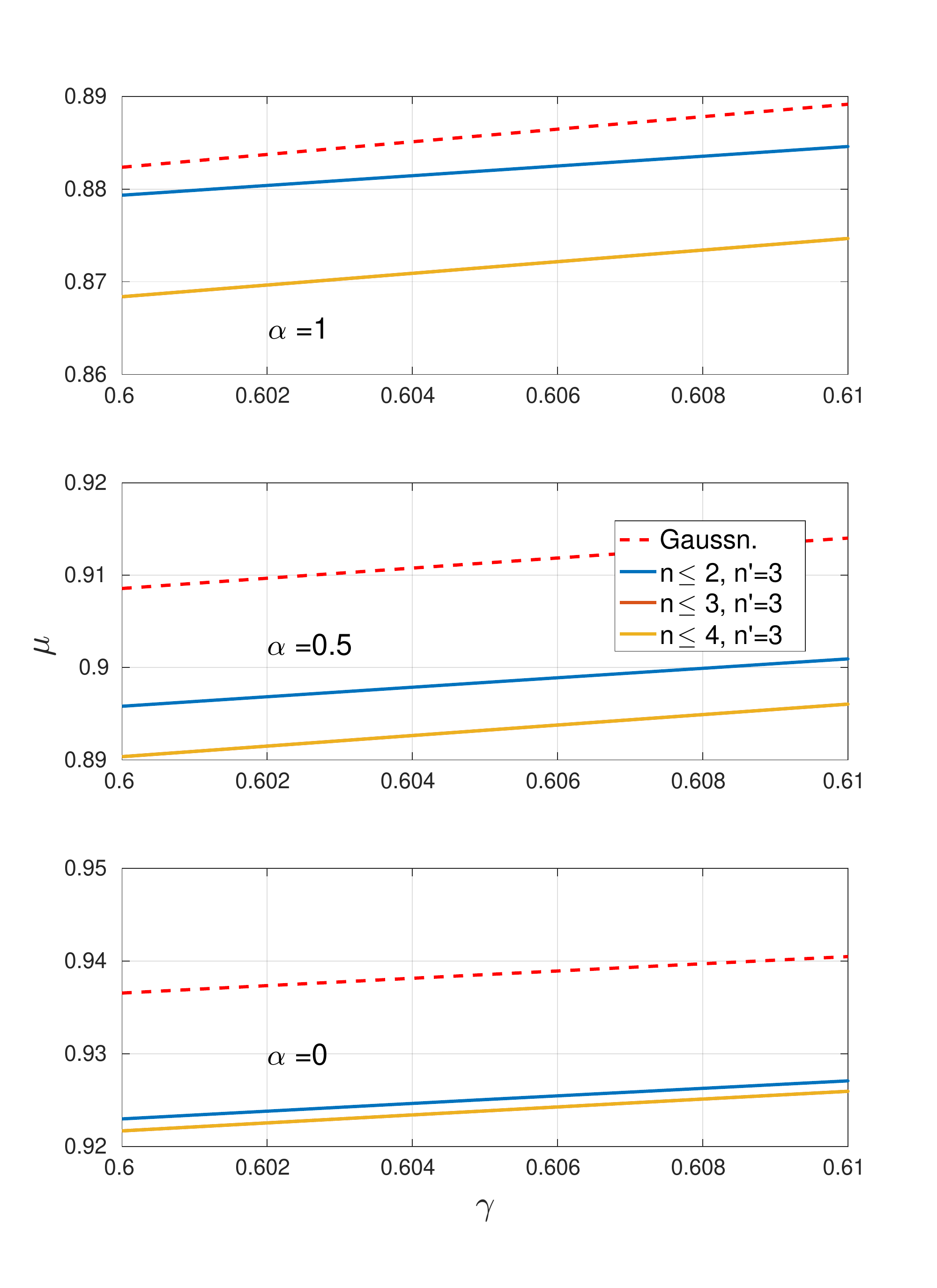}
\caption{\label{En2} Steerable states according to  inequality \eqref{ladeEuge} for different values of $\alpha$ in three scenarios where the number of observables of the steered party (Bob) is kept fixed ($n^{\prime}=3$) while the number of observables available to Alice is varied. Steering is possible in the region above the  limiting lines. Dashed line: steerability limit for  Gaussian measurements.  Notice that  for $n=3$ and $n=4$ the limiting curves coincide.} 
\end{figure}                                                                                                                             

Figure \ref{En2} corresponds to a scenario where $n^{\prime}$ (the number of setups available to the steered system) is kept fixed ($n^{\prime}=3$) while $n$ is varied. Inequality \eqref{ladeEuge} is used to reveal states steerable from Alice to Bob. As expected, due to the maximization on the LHS, adding new observables to the set  used by Alice cannot result in diminution of the ensemble of steerable states. As a matter of fact in this example $n=3$ and $n=4$ reveal the same set of steerable states. \\

\begin{figure}[htb]
\includegraphics[width=1\linewidth]{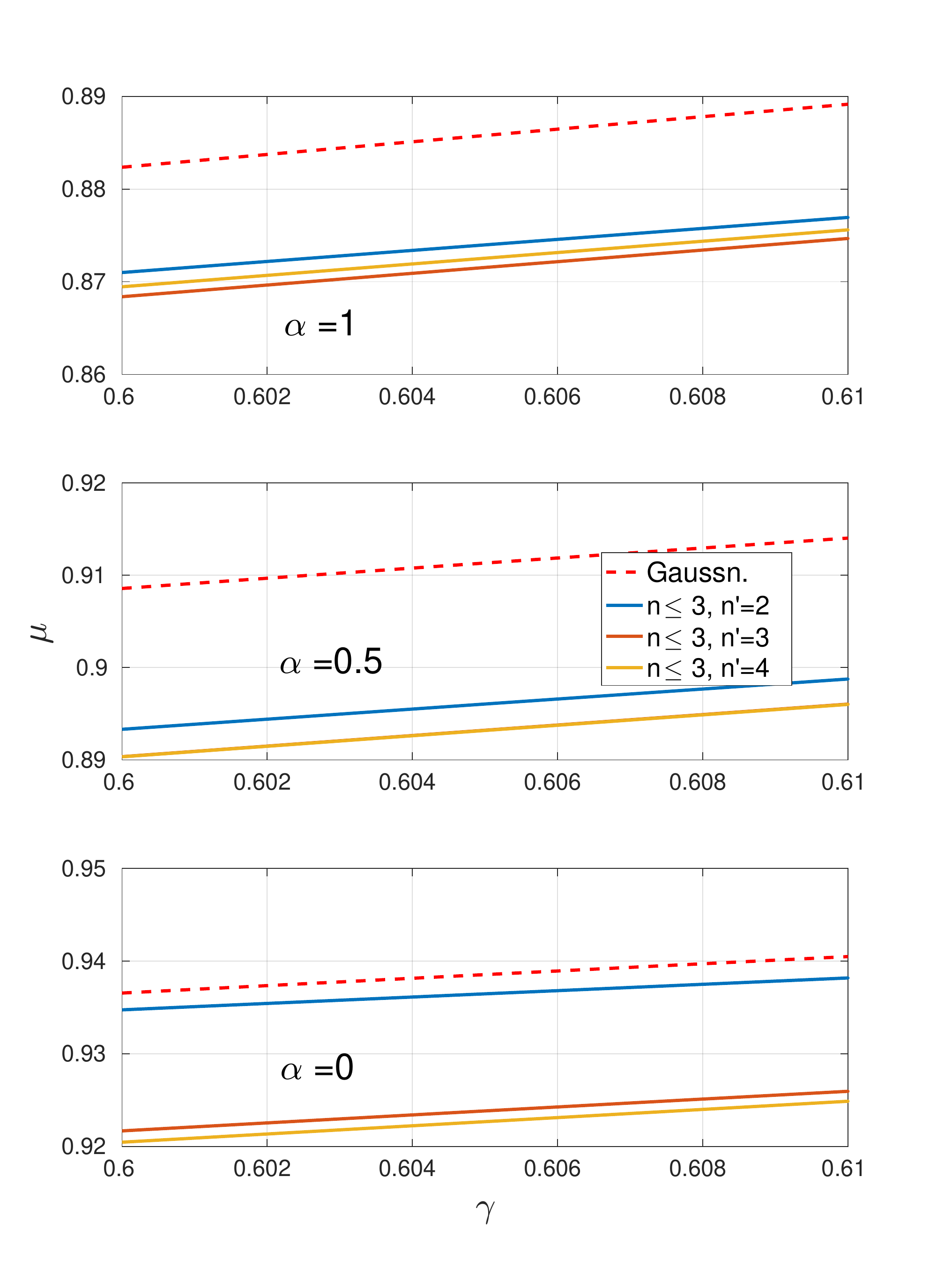}
\caption{\label{E2n} Steerable states according to  inequality \eqref{ladeEuge} for different values of $\alpha$ in three scenarios where the number of observables $n^{\prime}$ available to the steered party (Bob) is varied while the number of observables available to Alice is fixed ($n=3$).  
Steering is possible in the region above the  limiting lines. Dashed line: steerability limit for  Gaussian measurements. Notice that for $\alpha=0.5$ the red and yellow lines overlap. }
\end{figure}                                                                                                                             
The situation is different when $n$ (the number of Alice's observables) is kept fixed while $n^{\prime}$ is varied (Fig. \ref{E2n}). Notice that, depending on the state, the ability of inequality \eqref{ladeEuge} to reveal steerable states does not necessarily increase with $n^{\prime}$. This is due to the fact that a different inequality must be violated for each value of $n^{\prime}$.\\

\begin{figure}[htb]
\centering
\includegraphics[width=1\linewidth]{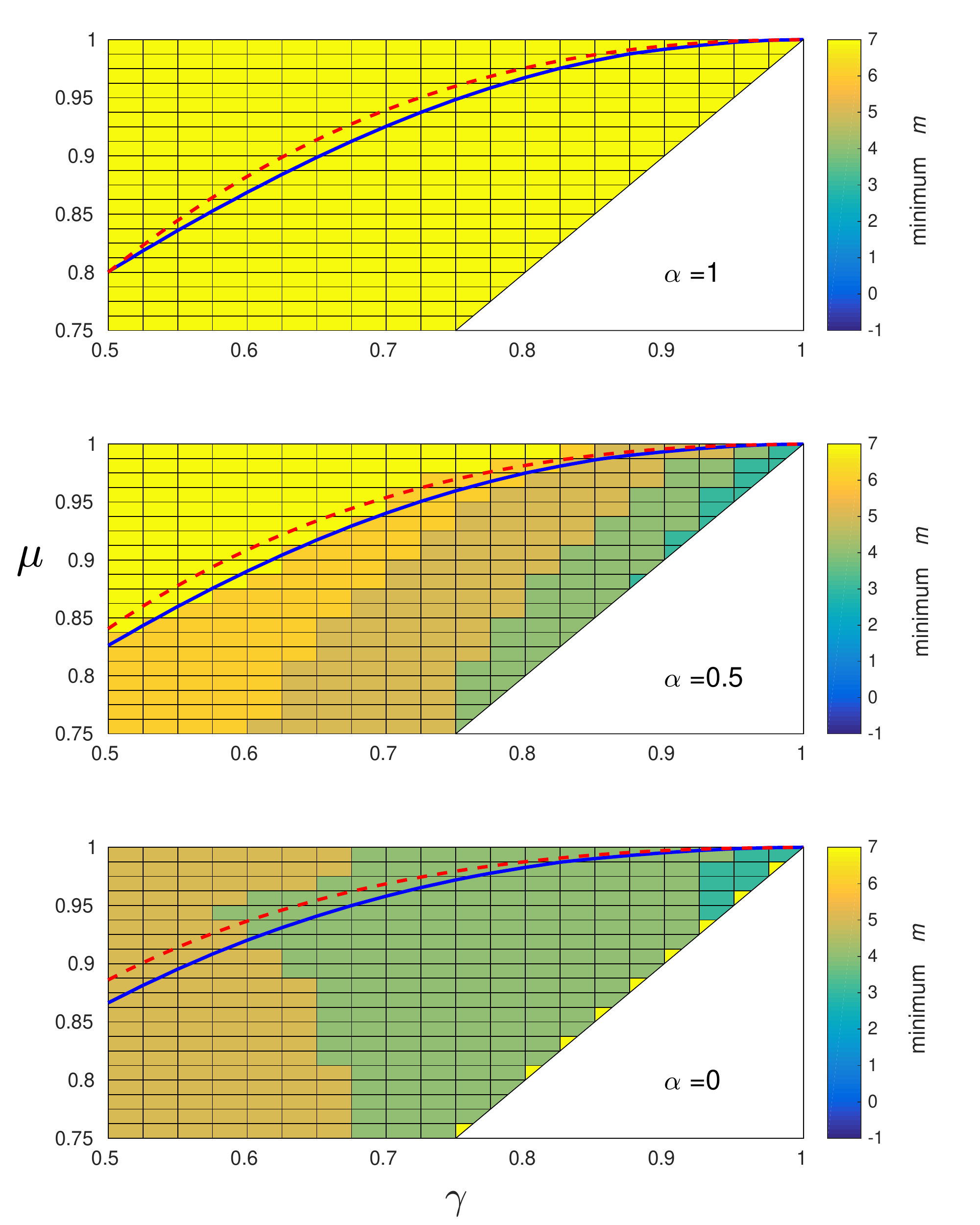}
\caption{\label{colorinche} Evaluation of inequality \eqref{ladeEuge} with $n^{\prime}=7$ for  symmetric states corresponding to the center of the colored cells. States steerable from Alice to Bob lie above the blue solid line.  The cell color indicates the minimum value $m \leq 7$ corresponding to Alice's observable set ${\lbrace A^{(m)}_i \rbrace}$ required to maximize the LHS of inequality \eqref{ladeEuge}. White zones correspond to separable states. The dashed red line indicates the steering limit for Gaussian measurements.} 
\end{figure}      

The previous results show that identification of steerable states by non-Gaussian measurements is sensitive to the symmetry of the setups available to both parties. We next address the question of whether asymmetric setups can be more ``efficient" than symmetrical ones. An ensemble containing a given number of observables will be considered  more efficient than another ensemble with a larger number of observables if it can reveal the same steerable states (same results with fewer resources).\\

Figure \ref{colorinche} shows the numerical evaluation of inequality  \eqref{ladeEuge} for $n^{\prime}=7$ with $n\leq 7$. The steerable states lie above the continuous curve. The minimum value of $m$ for which ${\lbrace A^{(m)}_i \rbrace}$ includes all the observables corresponding to the maximum of the LHS of \eqref{ladeEuge} is indicated by the background color. Notice that for $\alpha =0$, the use by Alice of observables contained in the set ${\lbrace A^{(4)}_i \rbrace}$ is sufficient to detect steering while Bob uses all observables in the ${\lbrace B^{(7)}_j \rbrace}$ set. In this case, the use of asymmetric setups can be considered as optimal in the sense that it provides the same steering identification than  larger symmetrical setups while involving fewer resources. 

This example shows that if the detection of steering is considered as a task, used for instance to certify entanglement between trusted and untrusted parties, for some symmetric states the use of asymmetric setups can be efficient and optimal.

\section{Conclusions}\label{sec:conclusions}

We have explored  quantum steering in symmetric two-mode Gaussian states. The systematic survey of symmetric states was facilitated by the introduction of a set of three independent parameters which allows mapping all symmetric Gaussian states onto the points of a three-dimensional cubic volume. Two different types of measurements were explored: Gaussian measurements and the discrete sets of non-Gaussian observables introduced in \citep{Ji2016}. Using the latter we have used three steering inequalities all of which provided examples showing that non-Gaussian measurements can outperform Gaussian measurements even for symmetric states.\\

Previous work concerned with Gaussian and non-Gaussian measurements has analyzed steering in asymmetric two-mode states. Here we have addressed for the first time the scenario, made possible by non-Gaussian measurements, where the global state is symmetrical while asymmetry arises from the use different number of observables by the two parties. We provide examples where asymmetric setups, involving a smaller number of resources on one party, can be as efficient in revealing steering as the corresponding symmetric setups with equal resources on both sides.\\

Our results provides new insight into the topic of non-Gaussian measurements applied to continuous variables systems. They were based on a specific set of observables \citep{Ji2016} and on a specific kind of steering inequalities \citep{Cavalcanti2007}. For completeness we mention that we have explored a second set of non-Gaussian observables introduced in \citep{Xiang17} where it was shown that these observables are suitable for revealing steering in non-symmetric states. However, we found that they do not outperform Gaussian measurements for symmetric states. 

The question of whether other types of non-Gaussian observables and steering criteria can further increase the identification of steerable states remains largely open and worth exploring.\\

\appendix*
\section{Equivalence of the two parameterizations}

We start by noticing that the state described by \eqref{standardformu} can be obtained by sending through a symmetric BS the state:
\begin{eqnarray}\label{bsinput}
V_{in}=Diag[u(p+m),\frac{p+n}{u},u(p-m),\frac{p-n}{u}]
\end{eqnarray}

We equate each term of the two expressions of the covariance matrices \eqref{bsinput} and \eqref{estadoparametrizado} corresponding to the BS input. Taking the logarithm on both sides of each equation and after some mathematical manipulations we obtain:
\begin{subequations}
\begin{eqnarray}\label{despejando}
\mu &=& \frac{1}{\sqrt{(p+m)(p+n)}} \\
\alpha &=& \dfrac{\log[(p-m)(p-n)]}{\log[(p+m)(p+n)]}\\
\gamma &=& \left[ \dfrac{(p+n)(p-m)}{(p-n)(p+m)}\right]^{\frac{1}{2(\alpha+1)}} \\
u &=& \left[ \left( \frac{p-n}{p-m}\right) \left( \frac{p+n}{p+m}\right)^{\alpha}\right] ^{\frac{1}{2(\alpha+1)}} 
\end{eqnarray}
\end{subequations}

Any physical covariance matrix of the form \eqref{standardformu} must verify: $p  \geq 1$, $p \geq \vert m \vert $, $p \geq \vert n \vert $,  $(p+m)(p+n) \geq 1$ and $(p-m)(p-n) \geq 1$ ( see \eqref{bsinput}). In consequence: $\mu \leq 1$, $\alpha \geq 0$, $\gamma > 0$ and $u > 0$. 

Also, with no loss of generality $m$ and $n$ can be chosen to satisfy  $m \geq \vert n \vert$  \citep{Marian08} in which case $0 < \gamma \leq 1$ and $0 < \alpha \leq 1$.\\

We conclude that any Gaussian symmetric state corresponding to the parameters $p, m, n$ in \eqref{standardformu} can be represented by a point inside the cube of unit edge in the $\gamma, \mu, \alpha$ space. The reciprocal statement is ensured by the fact that \eqref{estadoparametrizado} always  describes a physical state.\\

Finally, from the equality between \eqref{bsinput} and \eqref{estadoparametrizado} we obtain:
\begin{eqnarray}\label{paraentrelazamiento}
\left(\frac{\gamma}{\mu}\right)^{\alpha+1}&=& (p-m)(p+n)
\end{eqnarray}
In consequence $\gamma < \mu$ implies $(p-m)(p+n) < 1$ which is a sufficient condition for entanglement \citep{Marian08}.


\begin{thebibliography}{35}%
\makeatletter
\providecommand \@ifxundefined [1]{%
 \@ifx{#1\undefined}
}%
\providecommand \@ifnum [1]{%
 \ifnum #1\expandafter \@firstoftwo
 \else \expandafter \@secondoftwo
 \fi
}%
\providecommand \@ifx [1]{%
 \ifx #1\expandafter \@firstoftwo
 \else \expandafter \@secondoftwo
 \fi
}%
\providecommand \natexlab [1]{#1}%
\providecommand \enquote  [1]{``#1''}%
\providecommand \bibnamefont  [1]{#1}%
\providecommand \bibfnamefont [1]{#1}%
\providecommand \citenamefont [1]{#1}%
\providecommand \href@noop [0]{\@secondoftwo}%
\providecommand \href [0]{\begingroup \@sanitize@url \@href}%
\providecommand \@href[1]{\@@startlink{#1}\@@href}%
\providecommand \@@href[1]{\endgroup#1\@@endlink}%
\providecommand \@sanitize@url [0]{\catcode `\\12\catcode `\$12\catcode
  `\&12\catcode `\#12\catcode `\^12\catcode `\_12\catcode `\%12\relax}%
\providecommand \@@startlink[1]{}%
\providecommand \@@endlink[0]{}%
\providecommand \url  [0]{\begingroup\@sanitize@url \@url }%
\providecommand \@url [1]{\endgroup\@href {#1}{\urlprefix }}%
\providecommand \urlprefix  [0]{URL }%
\providecommand \Eprint [0]{\href }%
\providecommand \doibase [0]{http://dx.doi.org/}%
\providecommand \selectlanguage [0]{\@gobble}%
\providecommand \bibinfo  [0]{\@secondoftwo}%
\providecommand \bibfield  [0]{\@secondoftwo}%
\providecommand \translation [1]{[#1]}%
\providecommand \BibitemOpen [0]{}%
\providecommand \bibitemStop [0]{}%
\providecommand \bibitemNoStop [0]{.\EOS\space}%
\providecommand \EOS [0]{\spacefactor3000\relax}%
\providecommand \BibitemShut  [1]{\csname bibitem#1\endcsname}%
\let\auto@bib@innerbib\@empty
\bibitem [{\citenamefont {Ji}\ \emph {et~al.}(2016{\natexlab{a}})\citenamefont
  {Ji}, \citenamefont {Ji}, \citenamefont {Park},\ and\ \citenamefont
  {Nha}}]{Ji2016}%
  \BibitemOpen
  \bibfield  {author} {\bibinfo {author} {\bibfnamefont {S.-W.}\ \bibnamefont
  {Ji}}, \bibinfo {author} {\bibfnamefont {S.-W.}\ \bibnamefont {Ji}}, \bibinfo
  {author} {\bibfnamefont {J.}~\bibnamefont {Park}}, \ and\ \bibinfo {author}
  {\bibfnamefont {H.}~\bibnamefont {Nha}},\ }\href {\doibase
  https://doi.org/10.1038/srep29729} {\bibfield  {journal} {\bibinfo  {journal}
  {Scientific Reports}\ }\textbf {\bibinfo {volume} {6}} (\bibinfo {year}
  {2016}{\natexlab{a}}),\ https://doi.org/10.1038/srep29729}\BibitemShut
  {NoStop}%
\bibitem [{\citenamefont {Einstein}\ \emph {et~al.}(1935)\citenamefont
  {Einstein}, \citenamefont {Podolsky},\ and\ \citenamefont
  {Rosen}}]{Einstein1935}%
  \BibitemOpen
  \bibfield  {author} {\bibinfo {author} {\bibfnamefont {A.}~\bibnamefont
  {Einstein}}, \bibinfo {author} {\bibfnamefont {B.}~\bibnamefont {Podolsky}},
  \ and\ \bibinfo {author} {\bibfnamefont {N.}~\bibnamefont {Rosen}},\ }\href
  {\doibase 10.1103/PhysRev.47.777} {\bibfield  {journal} {\bibinfo  {journal}
  {Physical Review}\ }\textbf {\bibinfo {volume} {47}},\ \bibinfo {pages} {777}
  (\bibinfo {year} {1935})}\BibitemShut {NoStop}%
\bibitem [{\citenamefont {Quintino}\ \emph {et~al.}(2015)\citenamefont
  {Quintino}, \citenamefont {V\'ertesi}, \citenamefont {Cavalcanti},
  \citenamefont {Augusiak}, \citenamefont {Demianowicz}, \citenamefont
  {Ac\'{\i}n},\ and\ \citenamefont {Brunner}}]{Quintino2015}%
  \BibitemOpen
  \bibfield  {author} {\bibinfo {author} {\bibfnamefont {M.~T.}\ \bibnamefont
  {Quintino}}, \bibinfo {author} {\bibfnamefont {T.}~\bibnamefont {V\'ertesi}},
  \bibinfo {author} {\bibfnamefont {D.}~\bibnamefont {Cavalcanti}}, \bibinfo
  {author} {\bibfnamefont {R.}~\bibnamefont {Augusiak}}, \bibinfo {author}
  {\bibfnamefont {M.}~\bibnamefont {Demianowicz}}, \bibinfo {author}
  {\bibfnamefont {A.}~\bibnamefont {Ac\'{\i}n}}, \ and\ \bibinfo {author}
  {\bibfnamefont {N.}~\bibnamefont {Brunner}},\ }\href {\doibase
  10.1103/PhysRevA.92.032107} {\bibfield  {journal} {\bibinfo  {journal} {Phys.
  Rev. A}\ }\textbf {\bibinfo {volume} {92}},\ \bibinfo {pages} {032107}
  (\bibinfo {year} {2015})}\BibitemShut {NoStop}%
\bibitem [{\citenamefont {Schr{\"o}dinger}(1935)}]{Schrodinger1935}%
  \BibitemOpen
  \bibfield  {author} {\bibinfo {author} {\bibfnamefont {E.}~\bibnamefont
  {Schr{\"o}dinger}},\ }\href {\doibase 10.1017/S0305004100013554} {\bibfield
  {journal} {\bibinfo  {journal} {Mathematical Proceedings of the Cambridge
  Philosophical Society}\ }\textbf {\bibinfo {volume} {31}},\ \bibinfo {pages}
  {555–563} (\bibinfo {year} {1935})}\BibitemShut {NoStop}%
\bibitem [{\citenamefont {Gisin}(1991)}]{Gisin1991}%
  \BibitemOpen
  \bibfield  {author} {\bibinfo {author} {\bibfnamefont {N.}~\bibnamefont
  {Gisin}},\ }\href {\doibase https://doi.org/10.1016/0375-9601(91)90805-I}
  {\bibfield  {journal} {\bibinfo  {journal} {Physics Letters A}\ }\textbf
  {\bibinfo {volume} {154}},\ \bibinfo {pages} {201} (\bibinfo {year}
  {1991})}\BibitemShut {NoStop}%
\bibitem [{\citenamefont {Wiseman}\ \emph {et~al.}(2007)\citenamefont
  {Wiseman}, \citenamefont {Jones},\ and\ \citenamefont {Doherty}}]{Wiseman07}%
  \BibitemOpen
  \bibfield  {author} {\bibinfo {author} {\bibfnamefont {H.~M.}\ \bibnamefont
  {Wiseman}}, \bibinfo {author} {\bibfnamefont {S.~J.}\ \bibnamefont {Jones}},
  \ and\ \bibinfo {author} {\bibfnamefont {A.~C.}\ \bibnamefont {Doherty}},\
  }\href@noop {} {\bibfield  {journal} {\bibinfo  {journal} {Physical review
  letters}\ }\textbf {\bibinfo {volume} {98}},\ \bibinfo {pages} {140402}
  (\bibinfo {year} {2007})}\BibitemShut {NoStop}%
\bibitem [{\citenamefont {Werner}(2014)}]{Werner14}%
  \BibitemOpen
  \bibfield  {author} {\bibinfo {author} {\bibfnamefont {R.~F.}\ \bibnamefont
  {Werner}},\ }\href@noop {} {\bibfield  {journal} {\bibinfo  {journal}
  {Journal of Physics A: Mathematical and Theoretical}\ }\textbf {\bibinfo
  {volume} {47}},\ \bibinfo {pages} {424008} (\bibinfo {year}
  {2014})}\BibitemShut {NoStop}%
\bibitem [{\citenamefont {Branciard}\ \emph {et~al.}(2012)\citenamefont
  {Branciard}, \citenamefont {Cavalcanti}, \citenamefont {Walborn},
  \citenamefont {Scarani},\ and\ \citenamefont {Wiseman}}]{Branciard12}%
  \BibitemOpen
  \bibfield  {author} {\bibinfo {author} {\bibfnamefont {C.}~\bibnamefont
  {Branciard}}, \bibinfo {author} {\bibfnamefont {E.~G.}\ \bibnamefont
  {Cavalcanti}}, \bibinfo {author} {\bibfnamefont {S.~P.}\ \bibnamefont
  {Walborn}}, \bibinfo {author} {\bibfnamefont {V.}~\bibnamefont {Scarani}}, \
  and\ \bibinfo {author} {\bibfnamefont {H.~M.}\ \bibnamefont {Wiseman}},\
  }\href {\doibase 10.1103/PhysRevA.85.010301} {\bibfield  {journal} {\bibinfo
  {journal} {Phys. Rev. A}\ }\textbf {\bibinfo {volume} {85}},\ \bibinfo
  {pages} {010301} (\bibinfo {year} {2012})}\BibitemShut {NoStop}%
\bibitem [{\citenamefont {Cavalcanti}\ \emph {et~al.}(2013)\citenamefont
  {Cavalcanti}, \citenamefont {Hall},\ and\ \citenamefont
  {Wiseman}}]{Cavalcanti13}%
  \BibitemOpen
  \bibfield  {author} {\bibinfo {author} {\bibfnamefont {E.~G.}\ \bibnamefont
  {Cavalcanti}}, \bibinfo {author} {\bibfnamefont {M.~J.~W.}\ \bibnamefont
  {Hall}}, \ and\ \bibinfo {author} {\bibfnamefont {H.~M.}\ \bibnamefont
  {Wiseman}},\ }\href {\doibase 10.1103/PhysRevA.87.032306} {\bibfield
  {journal} {\bibinfo  {journal} {Phys. Rev. A}\ }\textbf {\bibinfo {volume}
  {87}},\ \bibinfo {pages} {032306} (\bibinfo {year} {2013})}\BibitemShut
  {NoStop}%
\bibitem [{\citenamefont {He}\ \emph {et~al.}(2015)\citenamefont {He},
  \citenamefont {Rosales-Z\'arate}, \citenamefont {Adesso},\ and\ \citenamefont
  {Reid}}]{CVteleportSteering}%
  \BibitemOpen
  \bibfield  {author} {\bibinfo {author} {\bibfnamefont {Q.}~\bibnamefont
  {He}}, \bibinfo {author} {\bibfnamefont {L.}~\bibnamefont
  {Rosales-Z\'arate}}, \bibinfo {author} {\bibfnamefont {G.}~\bibnamefont
  {Adesso}}, \ and\ \bibinfo {author} {\bibfnamefont {M.~D.}\ \bibnamefont
  {Reid}},\ }\href {\doibase 10.1103/PhysRevLett.115.180502} {\bibfield
  {journal} {\bibinfo  {journal} {Phys. Rev. Lett.}\ }\textbf {\bibinfo
  {volume} {115}},\ \bibinfo {pages} {180502} (\bibinfo {year}
  {2015})}\BibitemShut {NoStop}%
\bibitem [{\citenamefont {Passaro}\ \emph {et~al.}(2015)\citenamefont
  {Passaro}, \citenamefont {Cavalcanti}, \citenamefont {Skrzypczyk},\ and\
  \citenamefont {Ac{\'{\i}}n}}]{Passaro_2015}%
  \BibitemOpen
  \bibfield  {author} {\bibinfo {author} {\bibfnamefont {E.}~\bibnamefont
  {Passaro}}, \bibinfo {author} {\bibfnamefont {D.}~\bibnamefont {Cavalcanti}},
  \bibinfo {author} {\bibfnamefont {P.}~\bibnamefont {Skrzypczyk}}, \ and\
  \bibinfo {author} {\bibfnamefont {A.}~\bibnamefont {Ac{\'{\i}}n}},\ }\href
  {\doibase 10.1088/1367-2630/17/11/113010} {\bibfield  {journal} {\bibinfo
  {journal} {New Journal of Physics}\ }\textbf {\bibinfo {volume} {17}},\
  \bibinfo {pages} {113010} (\bibinfo {year} {2015})}\BibitemShut {NoStop}%
\bibitem [{\citenamefont {He}\ and\ \citenamefont
  {Reid}(2013)}]{genuinmultisteering}%
  \BibitemOpen
  \bibfield  {author} {\bibinfo {author} {\bibfnamefont {Q.~Y.}\ \bibnamefont
  {He}}\ and\ \bibinfo {author} {\bibfnamefont {M.~D.}\ \bibnamefont {Reid}},\
  }\href {\doibase 10.1103/PhysRevLett.111.250403} {\bibfield  {journal}
  {\bibinfo  {journal} {Phys. Rev. Lett.}\ }\textbf {\bibinfo {volume} {111}},\
  \bibinfo {pages} {250403} (\bibinfo {year} {2013})}\BibitemShut {NoStop}%
\bibitem [{\citenamefont {Jia}\ \emph {et~al.}(2020)\citenamefont {Jia},
  \citenamefont {Zhai}, \citenamefont {Yu}, \citenamefont {Wu},\ and\
  \citenamefont {Guo}}]{hierarchymultipartite}%
  \BibitemOpen
  \bibfield  {author} {\bibinfo {author} {\bibfnamefont {Z.-A.}\ \bibnamefont
  {Jia}}, \bibinfo {author} {\bibfnamefont {R.}~\bibnamefont {Zhai}}, \bibinfo
  {author} {\bibfnamefont {S.}~\bibnamefont {Yu}}, \bibinfo {author}
  {\bibfnamefont {Y.-C.}\ \bibnamefont {Wu}}, \ and\ \bibinfo {author}
  {\bibfnamefont {G.-C.}\ \bibnamefont {Guo}},\ }\href {\doibase
  10.1007/s11128-020-02922-z} {\bibfield  {journal} {\bibinfo  {journal}
  {Quantum Information Processing}\ }\textbf {\bibinfo {volume} {19}},\
  \bibinfo {pages} {419} (\bibinfo {year} {2020})}\BibitemShut {NoStop}%
\bibitem [{\citenamefont {Armstrong}\ \emph {et~al.}(2015)\citenamefont
  {Armstrong}, \citenamefont {Wang}, \citenamefont {Teh}, \citenamefont {Gong},
  \citenamefont {He}, \citenamefont {Janousek}, \citenamefont {Bachor},
  \citenamefont {Reid},\ and\ \citenamefont {Lam}}]{3partisteeringopticalnet}%
  \BibitemOpen
  \bibfield  {author} {\bibinfo {author} {\bibfnamefont {S.}~\bibnamefont
  {Armstrong}}, \bibinfo {author} {\bibfnamefont {M.}~\bibnamefont {Wang}},
  \bibinfo {author} {\bibfnamefont {R.~Y.}\ \bibnamefont {Teh}}, \bibinfo
  {author} {\bibfnamefont {Q.}~\bibnamefont {Gong}}, \bibinfo {author}
  {\bibfnamefont {Q.}~\bibnamefont {He}}, \bibinfo {author} {\bibfnamefont
  {J.}~\bibnamefont {Janousek}}, \bibinfo {author} {\bibfnamefont {H.-A.}\
  \bibnamefont {Bachor}}, \bibinfo {author} {\bibfnamefont {M.~D.}\
  \bibnamefont {Reid}}, \ and\ \bibinfo {author} {\bibfnamefont {P.~K.}\
  \bibnamefont {Lam}},\ }\href {\doibase 10.1038/nphys3202} {\bibfield
  {journal} {\bibinfo  {journal} {Nature Physics}\ }\textbf {\bibinfo {volume}
  {11}},\ \bibinfo {pages} {167} (\bibinfo {year} {2015})}\BibitemShut
  {NoStop}%
\bibitem [{\citenamefont {Cai}\ \emph {et~al.}(2020)\citenamefont {Cai},
  \citenamefont {Xiang}, \citenamefont {Liu}, \citenamefont {He},\ and\
  \citenamefont {Treps}}]{multisteeringfreqcomb}%
  \BibitemOpen
  \bibfield  {author} {\bibinfo {author} {\bibfnamefont {Y.}~\bibnamefont
  {Cai}}, \bibinfo {author} {\bibfnamefont {Y.}~\bibnamefont {Xiang}}, \bibinfo
  {author} {\bibfnamefont {Y.}~\bibnamefont {Liu}}, \bibinfo {author}
  {\bibfnamefont {Q.}~\bibnamefont {He}}, \ and\ \bibinfo {author}
  {\bibfnamefont {N.}~\bibnamefont {Treps}},\ }\href {\doibase
  10.1103/PhysRevResearch.2.032046} {\bibfield  {journal} {\bibinfo  {journal}
  {Phys. Rev. Research}\ }\textbf {\bibinfo {volume} {2}},\ \bibinfo {pages}
  {032046} (\bibinfo {year} {2020})}\BibitemShut {NoStop}%
\bibitem [{\citenamefont {Duan}\ \emph {et~al.}(2000)\citenamefont {Duan},
  \citenamefont {Giedke}, \citenamefont {Cirac},\ and\ \citenamefont
  {Zoller}}]{Duan00}%
  \BibitemOpen
  \bibfield  {author} {\bibinfo {author} {\bibfnamefont {L.-M.}\ \bibnamefont
  {Duan}}, \bibinfo {author} {\bibfnamefont {G.}~\bibnamefont {Giedke}},
  \bibinfo {author} {\bibfnamefont {J.~I.}\ \bibnamefont {Cirac}}, \ and\
  \bibinfo {author} {\bibfnamefont {P.}~\bibnamefont {Zoller}},\ }\href@noop {}
  {\bibfield  {journal} {\bibinfo  {journal} {Physical Review Letters}\
  }\textbf {\bibinfo {volume} {84}},\ \bibinfo {pages} {2722} (\bibinfo {year}
  {2000})}\BibitemShut {NoStop}%
\bibitem [{\citenamefont {Adesso}\ \emph {et~al.}(2014)\citenamefont {Adesso},
  \citenamefont {Ragy},\ and\ \citenamefont {Lee}}]{Adesso14}%
  \BibitemOpen
  \bibfield  {author} {\bibinfo {author} {\bibfnamefont {G.}~\bibnamefont
  {Adesso}}, \bibinfo {author} {\bibfnamefont {S.}~\bibnamefont {Ragy}}, \ and\
  \bibinfo {author} {\bibfnamefont {A.~R.}\ \bibnamefont {Lee}},\ }\href@noop
  {} {\bibfield  {journal} {\bibinfo  {journal} {Open Systems \& Information
  Dynamics}\ }\textbf {\bibinfo {volume} {21}},\ \bibinfo {pages} {1440001}
  (\bibinfo {year} {2014})}\BibitemShut {NoStop}%
\bibitem [{\citenamefont {Marian}\ and\ \citenamefont
  {Marian}(2008{\natexlab{a}})}]{Marian08}%
  \BibitemOpen
  \bibfield  {author} {\bibinfo {author} {\bibfnamefont {P.}~\bibnamefont
  {Marian}}\ and\ \bibinfo {author} {\bibfnamefont {T.~A.}\ \bibnamefont
  {Marian}},\ }\href@noop {} {\bibfield  {journal} {\bibinfo  {journal} {The
  European Physical Journal Special Topics}\ }\textbf {\bibinfo {volume}
  {160}},\ \bibinfo {pages} {281} (\bibinfo {year}
  {2008}{\natexlab{a}})}\BibitemShut {NoStop}%
\bibitem [{\citenamefont {Kim}\ \emph {et~al.}(2002)\citenamefont {Kim},
  \citenamefont {Son}, \citenamefont {Bu\ifmmode~\check{z}\else \v{z}\fi{}ek},\
  and\ \citenamefont {Knight}}]{Kim2002}%
  \BibitemOpen
  \bibfield  {author} {\bibinfo {author} {\bibfnamefont {M.~S.}\ \bibnamefont
  {Kim}}, \bibinfo {author} {\bibfnamefont {W.}~\bibnamefont {Son}}, \bibinfo
  {author} {\bibfnamefont {V.}~\bibnamefont {Bu\ifmmode~\check{z}\else
  \v{z}\fi{}ek}}, \ and\ \bibinfo {author} {\bibfnamefont {P.~L.}\ \bibnamefont
  {Knight}},\ }\href {\doibase 10.1103/PhysRevA.65.032323} {\bibfield
  {journal} {\bibinfo  {journal} {Phys. Rev. A}\ }\textbf {\bibinfo {volume}
  {65}},\ \bibinfo {pages} {032323} (\bibinfo {year} {2002})}\BibitemShut
  {NoStop}%
\bibitem [{\citenamefont {Wolf}\ \emph {et~al.}(2003)\citenamefont {Wolf},
  \citenamefont {Eisert},\ and\ \citenamefont {Plenio}}]{Wolf2003}%
  \BibitemOpen
  \bibfield  {author} {\bibinfo {author} {\bibfnamefont {M.~M.}\ \bibnamefont
  {Wolf}}, \bibinfo {author} {\bibfnamefont {J.}~\bibnamefont {Eisert}}, \ and\
  \bibinfo {author} {\bibfnamefont {M.~B.}\ \bibnamefont {Plenio}},\ }\href
  {\doibase 10.1103/PhysRevLett.90.047904} {\bibfield  {journal} {\bibinfo
  {journal} {Phys. Rev. Lett.}\ }\textbf {\bibinfo {volume} {90}},\ \bibinfo
  {pages} {047904} (\bibinfo {year} {2003})}\BibitemShut {NoStop}%
\bibitem [{\citenamefont {Laurat}\ \emph {et~al.}(2005)\citenamefont {Laurat},
  \citenamefont {Keller}, \citenamefont {Oliveira-Huguenin}, \citenamefont
  {Fabre}, \citenamefont {Coudreau}, \citenamefont {Serafini}, \citenamefont
  {Adesso},\ and\ \citenamefont {Illuminati}}]{Laurat_2005}%
  \BibitemOpen
  \bibfield  {author} {\bibinfo {author} {\bibfnamefont {J.}~\bibnamefont
  {Laurat}}, \bibinfo {author} {\bibfnamefont {G.}~\bibnamefont {Keller}},
  \bibinfo {author} {\bibfnamefont {J.~A.}\ \bibnamefont {Oliveira-Huguenin}},
  \bibinfo {author} {\bibfnamefont {C.}~\bibnamefont {Fabre}}, \bibinfo
  {author} {\bibfnamefont {T.}~\bibnamefont {Coudreau}}, \bibinfo {author}
  {\bibfnamefont {A.}~\bibnamefont {Serafini}}, \bibinfo {author}
  {\bibfnamefont {G.}~\bibnamefont {Adesso}}, \ and\ \bibinfo {author}
  {\bibfnamefont {F.}~\bibnamefont {Illuminati}},\ }\href {\doibase
  10.1088/1464-4266/7/12/021} {\bibfield  {journal} {\bibinfo  {journal}
  {Journal of Optics B: Quantum and Semiclassical Optics}\ }\textbf {\bibinfo
  {volume} {7}},\ \bibinfo {pages} {S577} (\bibinfo {year} {2005})}\BibitemShut
  {NoStop}%
\bibitem [{\citenamefont {Xiang}\ \emph {et~al.}(2011)\citenamefont {Xiang},
  \citenamefont {Song}, \citenamefont {Wen},\ and\ \citenamefont
  {Shi}}]{Xiangsqueezed}%
  \BibitemOpen
  \bibfield  {author} {\bibinfo {author} {\bibfnamefont {S.~H.}\ \bibnamefont
  {Xiang}}, \bibinfo {author} {\bibfnamefont {K.~H.}\ \bibnamefont {Song}},
  \bibinfo {author} {\bibfnamefont {W.}~\bibnamefont {Wen}}, \ and\ \bibinfo
  {author} {\bibfnamefont {Z.~G.}\ \bibnamefont {Shi}},\ }\href {\doibase
  10.1140/epjd/e2011-10546-1} {\bibfield  {journal} {\bibinfo  {journal} {The
  European Physical Journal D}\ }\textbf {\bibinfo {volume} {62}},\ \bibinfo
  {pages} {289} (\bibinfo {year} {2011})}\BibitemShut {NoStop}%
\bibitem [{\citenamefont {Marian}\ and\ \citenamefont
  {Marian}(2008{\natexlab{b}})}]{MarianSymmetricG}%
  \BibitemOpen
  \bibfield  {author} {\bibinfo {author} {\bibfnamefont {P.}~\bibnamefont
  {Marian}}\ and\ \bibinfo {author} {\bibfnamefont {T.~A.}\ \bibnamefont
  {Marian}},\ }\href {\doibase 10.1140/epjst/e2008-00731-x} {\bibfield
  {journal} {\bibinfo  {journal} {The European Physical Journal Special
  Topics}\ }\textbf {\bibinfo {volume} {160}},\ \bibinfo {pages} {281}
  (\bibinfo {year} {2008}{\natexlab{b}})}\BibitemShut {NoStop}%
\bibitem [{\citenamefont {Chen}\ and\ \citenamefont {Qiu}(2003)}]{CHEN2003191}%
  \BibitemOpen
  \bibfield  {author} {\bibinfo {author} {\bibfnamefont {X.-Y.}\ \bibnamefont
  {Chen}}\ and\ \bibinfo {author} {\bibfnamefont {P.-L.}\ \bibnamefont {Qiu}},\
  }\href {\doibase https://doi.org/10.1016/S0375-9601(03)00916-2} {\bibfield
  {journal} {\bibinfo  {journal} {Physics Letters A}\ }\textbf {\bibinfo
  {volume} {314}},\ \bibinfo {pages} {191} (\bibinfo {year}
  {2003})}\BibitemShut {NoStop}%
\bibitem [{\citenamefont {Li}\ \emph {et~al.}(2016)\citenamefont {Li},
  \citenamefont {Xu}, \citenamefont {Yuan},\ and\ \citenamefont
  {Wang}}]{Li_2016}%
  \BibitemOpen
  \bibfield  {author} {\bibinfo {author} {\bibfnamefont {H.-M.}\ \bibnamefont
  {Li}}, \bibinfo {author} {\bibfnamefont {X.-X.}\ \bibnamefont {Xu}}, \bibinfo
  {author} {\bibfnamefont {H.-C.}\ \bibnamefont {Yuan}}, \ and\ \bibinfo
  {author} {\bibfnamefont {Z.}~\bibnamefont {Wang}},\ }\href {\doibase
  10.1088/1674-1056/25/10/104203} {\bibfield  {journal} {\bibinfo  {journal}
  {Chinese Physics B}\ }\textbf {\bibinfo {volume} {25}},\ \bibinfo {pages}
  {104203} (\bibinfo {year} {2016})}\BibitemShut {NoStop}%
\bibitem [{\citenamefont {Cuzminschi}\ \emph {et~al.}(2021)\citenamefont
  {Cuzminschi}, \citenamefont {Zubarev},\ and\ \citenamefont
  {Isar}}]{Cuzminschi2021}%
  \BibitemOpen
  \bibfield  {author} {\bibinfo {author} {\bibfnamefont {M.}~\bibnamefont
  {Cuzminschi}}, \bibinfo {author} {\bibfnamefont {A.}~\bibnamefont {Zubarev}},
  \ and\ \bibinfo {author} {\bibfnamefont {A.}~\bibnamefont {Isar}},\ }\href
  {\doibase https://doi.org/10.1038/s41598-021-03752-4} {\bibfield  {journal}
  {\bibinfo  {journal} {Scientific Reports}\ }\textbf {\bibinfo {volume}
  {11}},\ \bibinfo {pages} {24286} (\bibinfo {year} {2021})}\BibitemShut
  {NoStop}%
\bibitem [{\citenamefont {Kogias}\ \emph {et~al.}(2015)\citenamefont {Kogias},
  \citenamefont {Lee}, \citenamefont {Ragy},\ and\ \citenamefont
  {Adesso}}]{Kogias15}%
  \BibitemOpen
  \bibfield  {author} {\bibinfo {author} {\bibfnamefont {I.}~\bibnamefont
  {Kogias}}, \bibinfo {author} {\bibfnamefont {A.~R.}\ \bibnamefont {Lee}},
  \bibinfo {author} {\bibfnamefont {S.}~\bibnamefont {Ragy}}, \ and\ \bibinfo
  {author} {\bibfnamefont {G.}~\bibnamefont {Adesso}},\ }\href@noop {}
  {\bibfield  {journal} {\bibinfo  {journal} {Physical review letters}\
  }\textbf {\bibinfo {volume} {114}},\ \bibinfo {pages} {060403} (\bibinfo
  {year} {2015})}\BibitemShut {NoStop}%
\bibitem [{\citenamefont {Kogias}\ and\ \citenamefont
  {Adesso}(2015)}]{Kogias_15}%
  \BibitemOpen
  \bibfield  {author} {\bibinfo {author} {\bibfnamefont {I.}~\bibnamefont
  {Kogias}}\ and\ \bibinfo {author} {\bibfnamefont {G.}~\bibnamefont
  {Adesso}},\ }\href {\doibase 10.1364/JOSAB.32.000A27} {\bibfield  {journal}
  {\bibinfo  {journal} {J. Opt. Soc. Am. B}\ }\textbf {\bibinfo {volume}
  {32}},\ \bibinfo {pages} {A27} (\bibinfo {year} {2015})}\BibitemShut
  {NoStop}%
\bibitem [{\citenamefont {Tatham}\ and\ \citenamefont
  {Korolkova}(2014)}]{Tatham14}%
  \BibitemOpen
  \bibfield  {author} {\bibinfo {author} {\bibfnamefont {R.}~\bibnamefont
  {Tatham}}\ and\ \bibinfo {author} {\bibfnamefont {N.}~\bibnamefont
  {Korolkova}},\ }\href@noop {} {\bibfield  {journal} {\bibinfo  {journal}
  {Physical Review A}\ }\textbf {\bibinfo {volume} {89}},\ \bibinfo {pages}
  {012308} (\bibinfo {year} {2014})}\BibitemShut {NoStop}%
\bibitem [{\citenamefont {Ji}\ \emph {et~al.}(2016{\natexlab{b}})\citenamefont
  {Ji}, \citenamefont {Lee}, \citenamefont {Park},\ and\ \citenamefont
  {Nha}}]{Ji16}%
  \BibitemOpen
  \bibfield  {author} {\bibinfo {author} {\bibfnamefont {S.-W.}\ \bibnamefont
  {Ji}}, \bibinfo {author} {\bibfnamefont {J.}~\bibnamefont {Lee}}, \bibinfo
  {author} {\bibfnamefont {J.}~\bibnamefont {Park}}, \ and\ \bibinfo {author}
  {\bibfnamefont {H.}~\bibnamefont {Nha}},\ }\href@noop {} {\bibfield
  {journal} {\bibinfo  {journal} {Scientific reports}\ }\textbf {\bibinfo
  {volume} {6}},\ \bibinfo {pages} {29729} (\bibinfo {year}
  {2016}{\natexlab{b}})}\BibitemShut {NoStop}%
\bibitem [{\citenamefont {Reid}(1989)}]{Reid89}%
  \BibitemOpen
  \bibfield  {author} {\bibinfo {author} {\bibfnamefont {M.~D.}\ \bibnamefont
  {Reid}},\ }\href {\doibase 10.1103/PhysRevA.40.913} {\bibfield  {journal}
  {\bibinfo  {journal} {Phys. Rev. A}\ }\textbf {\bibinfo {volume} {40}},\
  \bibinfo {pages} {913} (\bibinfo {year} {1989})}\BibitemShut {NoStop}%
\bibitem [{\citenamefont {Cavalcanti}\ and\ \citenamefont
  {Reid}(2007)}]{Cavalcanti2007}%
  \BibitemOpen
  \bibfield  {author} {\bibinfo {author} {\bibfnamefont {E.~G.}\ \bibnamefont
  {Cavalcanti}}\ and\ \bibinfo {author} {\bibfnamefont {M.~D.}\ \bibnamefont
  {Reid}},\ }\href@noop {} {\bibfield  {journal} {\bibinfo  {journal} {Journal
  of Modern Optics}\ }\textbf {\bibinfo {volume} {54}},\ \bibinfo {pages}
  {2373} (\bibinfo {year} {2007})}\BibitemShut {NoStop}%
\bibitem [{\citenamefont {Cavalcanti}\ \emph {et~al.}(2009)\citenamefont
  {Cavalcanti}, \citenamefont {Jones}, \citenamefont {Wiseman},\ and\
  \citenamefont {Reid}}]{Cavalcanti09}%
  \BibitemOpen
  \bibfield  {author} {\bibinfo {author} {\bibfnamefont {E.~G.}\ \bibnamefont
  {Cavalcanti}}, \bibinfo {author} {\bibfnamefont {S.~J.}\ \bibnamefont
  {Jones}}, \bibinfo {author} {\bibfnamefont {H.~M.}\ \bibnamefont {Wiseman}},
  \ and\ \bibinfo {author} {\bibfnamefont {M.~D.}\ \bibnamefont {Reid}},\
  }\href@noop {} {\bibfield  {journal} {\bibinfo  {journal} {Physical Review
  A}\ }\textbf {\bibinfo {volume} {80}},\ \bibinfo {pages} {032112} (\bibinfo
  {year} {2009})}\BibitemShut {NoStop}%
\bibitem [{\citenamefont {Reid}\ \emph {et~al.}(2009)\citenamefont {Reid},
  \citenamefont {Drummond}, \citenamefont {Bowen}, \citenamefont {Cavalcanti},
  \citenamefont {Lam}, \citenamefont {Bachor}, \citenamefont {Andersen},\ and\
  \citenamefont {Leuchs}}]{Reid_2009}%
  \BibitemOpen
  \bibfield  {author} {\bibinfo {author} {\bibfnamefont {M.}~\bibnamefont
  {Reid}}, \bibinfo {author} {\bibfnamefont {P.}~\bibnamefont {Drummond}},
  \bibinfo {author} {\bibfnamefont {W.}~\bibnamefont {Bowen}}, \bibinfo
  {author} {\bibfnamefont {E.~G.}\ \bibnamefont {Cavalcanti}}, \bibinfo
  {author} {\bibfnamefont {P.~K.}\ \bibnamefont {Lam}}, \bibinfo {author}
  {\bibfnamefont {H.}~\bibnamefont {Bachor}}, \bibinfo {author} {\bibfnamefont
  {U.~L.}\ \bibnamefont {Andersen}}, \ and\ \bibinfo {author} {\bibfnamefont
  {G.}~\bibnamefont {Leuchs}},\ }\href@noop {} {\bibfield  {journal} {\bibinfo
  {journal} {Reviews of Modern Physics}\ }\textbf {\bibinfo {volume} {81}},\
  \bibinfo {pages} {1727} (\bibinfo {year} {2009})}\BibitemShut {NoStop}%
\bibitem [{\citenamefont {Xiang}\ \emph {et~al.}(2017)\citenamefont {Xiang},
  \citenamefont {Xu}, \citenamefont {Mi{\v{s}}ta~Jr}, \citenamefont
  {Tufarelli}, \citenamefont {He},\ and\ \citenamefont {Adesso}}]{Xiang17}%
  \BibitemOpen
  \bibfield  {author} {\bibinfo {author} {\bibfnamefont {Y.}~\bibnamefont
  {Xiang}}, \bibinfo {author} {\bibfnamefont {B.}~\bibnamefont {Xu}}, \bibinfo
  {author} {\bibfnamefont {L.}~\bibnamefont {Mi{\v{s}}ta~Jr}}, \bibinfo
  {author} {\bibfnamefont {T.}~\bibnamefont {Tufarelli}}, \bibinfo {author}
  {\bibfnamefont {Q.}~\bibnamefont {He}}, \ and\ \bibinfo {author}
  {\bibfnamefont {G.}~\bibnamefont {Adesso}},\ }\href@noop {} {\bibfield
  {journal} {\bibinfo  {journal} {Physical Review A}\ }\textbf {\bibinfo
  {volume} {96}},\ \bibinfo {pages} {042326} (\bibinfo {year}
  {2017})}\BibitemShut {NoStop}%
\end{thebibliography}

%

\end{document}